\RequirePackage{fix-cm}

\documentclass[twocolumn]{svjour3}          

\smartqed 

\usepackage{graphicx}
\usepackage{siunitx}
\usepackage{subfigure}

\newcommand{\Ba}{\textsuperscript{138}\text{Ba}\textsuperscript{+}\,}
\journalname{Applied Physics B}

\begin{document}

\title{Efficient photoionization for barium ion trapping using a dipole-allowed resonant two-photon transition}

\author{G. Leschhorn        \and
        T. Hasegawa            \and
        T. Schaetz 
}

\institute{G. Leschhorn \and
           T. Schaetz   \at
              Max-Planck-Institut for Quantum Optics \\
              Hans-Kopfermann-Str.1\\
              85748 Garching, Germany\\
              Tel.: +49 89-32905-199\\
              Fax: +49 89-32905-200\\
              \email{tobias.schaetz@mpq.mpg.de}          
					\and
					T. Schaetz	\at
							Albert-Ludwigs-Universit\"at Freiburg\\
							Hermann-Herder-Str. 3a\\
							79104 Freiburg, Germany\\
          \and
           T. Hasegawa  \at
              Faculty of Science and Technology,\\
              Department of Physics, Keio University\\
              3-14-1 Hiyoshi, Kohoku-ku\\
              Yokohama 223, Japan}

\date{}

\maketitle

\begin{abstract}
Two efficient and isotope-selective resonant two-photon ionization techniques for loading barium ions into radio-frequency (RF)-traps are demonstrated. The scheme of using the strong dipole-allowed transition $6s^2\, ^1\text{S}_0\rightarrow 6s6p\, ^1\text{P}_1$ at $\lambda=553\,\text{nm}$ as a first step towards ionization is compared to the established technique of using a weak intercombination line ($6s^2\, ^1\text{S}_0\rightarrow 5d6p\, ^3\text{D}_1,\,\lambda=413\,\text{nm}$). An increase of two orders of magnitude in the ionization efficiency is found favoring the transition at 553\,nm. This technique can be implemented using commercial all-solid-state laser systems and is expected to be advantageous compared to other narrowband photoionization schemes of barium in cases where highest efficiency and isotope-selectivity are required. 
\end{abstract}

\section{Introduction}
\label{introduction}
Ion traps have become an important tool in a growing number of fields, like precision measurements of fundamental constants\cite{rosenband08}, testing general relativity\cite{ChouScience10}, frequency standards\cite{chou10}, quantum information processing\cite{haeffner08,Home09}, quantum simulation\cite{Friedenauer08,Lanyon11} or mass spectrometry\cite{Paul90}. During the last decade, the state-of-the-art loading of ion traps has changed from electron beam bombardment to photoionization of neutral atoms inside the trapping region. The main reasons are a higher efficieny and the isotope selectivity of most of the photoionization schemes presented so far. Additional beneficial effects, like strongly reduced static charging of dielectrics\cite{Gulde01}, lower motional heating rates of the ions after loading due to the suppression of patch potentials\cite{Turchette00} and easier loading of surface-electrode traps with a substantially reduced trap depth\cite{Stick06} is of great advantage in the laboratories. Loading of Paul traps by narrowband cw photoionization\cite{Hurst79} has been achieved by several groups on a growing number of atomic species (Ba\cite{steele07,Wang11}, Mg\cite{kjaegaard00}, Ca\cite{Gulde01,lucas04,Tanaka05}, Yb\cite{Johanning11}, Sr\cite{Brownnutt07,vant06}, In\cite{phd_ludsteck} etc.). Several of these schemes have been applied to photoionization loading techniques that substitute the resistively heated atomic oven as source of neutral atoms. Instead, non-evaporating sources, for instance laser ablation targets \cite{Hendricks07} or magneto-optical traps\cite{cetina07} are used. These techniques additionally increase the efficiency and the advantages of photoionization loading and were mainly developed to load micro-structured surface traps. Furthermore, non-isotope selective loading schemes, e.g. using ultra-short pulses\cite{deslauriersmonroe06} or laser ablation\cite{leibrandt07} have been realized which share the advantage of being applicable to most laser-cooled ion species with a further increased efficiency. 

In this paper, a novel loading technique for barium by isotope-selective photoionization using a dipole-allowed two-photon transition is demonstrated. A substantial increase in ion loading efficiency compared to established schemes\cite{rotter08} is found. The presented technique promises to provide substantial advantages in experiments that depend on maximizing the efficiency or on a fast, controlled and minimal invasive reloading of ions.

\section{Experiment}
\label{experiment}
The ion trap used for the experiments presented in this paper consists of four cylindrical, gold plated copper rods of 2\,mm diameter for radial confinement and two additional ring shaped electrodes around the rods providing axial confinement. The rods are arranged in a quadrupole configuration with a minimal distance of 1.12\,mm between the trap center and the rod surface. An RF voltage at $\Omega_\text{RF}=2\pi\times6.8\,\text{MHz}$ is applied on the rod-electrodes. For \Ba the typically used radial secular frequency amounts to $2\pi\times470\,\text{kHz}$. The 5\,mm-thick ring electrodes that are centered around the trap axis are spaced by 15\,mm and have an outer diameter of 24\,mm and an inner diameter of 8\,mm. A DC voltage applied to these ring electrodes leads to a static, nearly harmonic confinement potential along the axis of the linear Paul trap. The trap setup is housed in an ultra-high vacuum chamber at a pressure of $2\times10^{-8}\,\text{Pa}$. Ions confined in the trap can be observed by imaging their fluorescence light during Doppler laser-cooling with a two-lens condenser on an electron-multiplying CCD camera. The magnification factor of the imaging system is approximately 10 and in conjunction with the trap parameters a single and up to approximately 50 individual ions that form a crystalline structure\cite{diedrich87b,Wineland87} can be observed.

\begin{figure}
	\centering
		\includegraphics[width=0.5\textwidth]{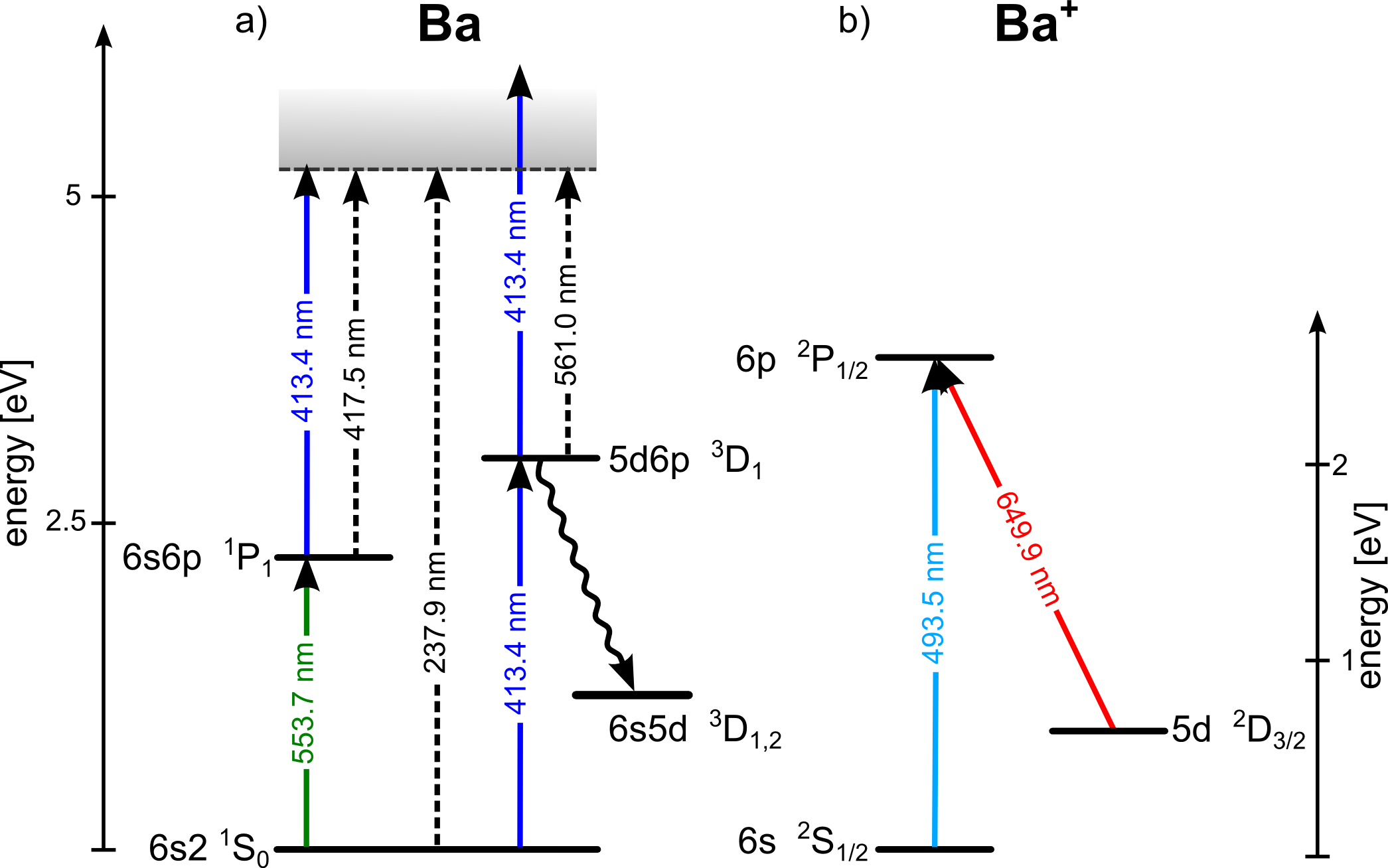}
	\caption{Energy level scheme of a) neutral and b) singly ionized barium. Only the relevant levels and transitions are shown. Solid lines indicate experimentally realized transition, whereas dashed lines are only given for comparison. For coarse orientation, an energy scale is given for both cases. Photoionization of neutral barium is experimentally realized using two different schemes, that both use one resonant excitation followed by a non-resonant excitation into the continuum. In scheme B, a 553.7 nm photon excites the $^1\text{P}_1$ state of $2\pi\times18.9\,\text{MHz}$ natural linewidth, corresponding to 8.4\,ns lifetime. Another photon at 413\,nm excites the atom above the ionization threshold at 5.21\,eV. The other ionization scheme (scheme A) works with two photons at 413\,nm. The first one couples the ground state with the $5d6p\, ^3\text{D}_1$ level. The second photon excites the electron deeply into the continuum. Note that in scheme A, the atoms are optically pumped (wavy line) to metastable states because the spontaneous emission rate to the states $6s5d\, ^3\text{D}_{1,2}$ ($3.8\times10^7\,\text{s}^{-1}$ to $^3\text{D}_1$, $1.9\times10^7\,\text{s}^{-1}$ to $^3\text{D}_2$) is larger than that to the ground state ($1.5\times10^6\,\text{s}^{-1}$). Laser cooling on the singly charged ion is done on the $\Lambda$-system spanned by the $^2\text{S}_{1/2}$, $^2\text{P}_{1/2}$ and $^2\text{D}_{3/2}$ levels, using photons near 493\,nm and near 650\,nm.}
	\label{fig:Ba_energy_level}
\end{figure}

Two different, resonant two-step excitations from the ground state $6s^2\, ^1\text{S}_0$ of neutral $^{138}\text{Ba}$ to the continuum are used to produce singly ionized barium. Since in both cases, the first step is realized by a resonant transition and the isotope shifts are sufficiently large compared to the laser linewidth and the Doppler and power broadening of the transition, both photoionization schemes are isotope selective. Figure \ref{fig:Ba_energy_level}a) gives a summary of the relevant transitions of neutral barium. The first scheme requires two photons at 413\,nm and is resonant with the weak intercombination line of the $6s^2\, ^1\text{S}_0\rightarrow 5d6p\, ^3\text{D}_1$ transition. It was first demonstrated in\cite{rotter08} and will be labeled scheme A in the following. Despite the weak transition strength, this scheme benefits from the fact that light of only one frequency has to be provided that can be generated by a commercial and robust diode laser. Within the framework of this paper, this scheme has been realized for comparison. The 413\,nm laser propagates on axis through the trap setup and is focused to a beam waist of $w_\text{413}=440$\,\micro m ($1/e^2$ of intensity) at the trap center.

The second scheme, labeled scheme B, is based on the strong dipole allowed $6s^2\, ^1\text{S}_0\rightarrow 6s6p\, ^1\text{P}_1$ transition at 553\,nm\cite{Armstrong93} to provide the first excitation. The second step from the $^1\text{P}_1$ state to the continuum requires a wavelength \textless 418\,nm. Therefore, a 413\,nm photon is sufficient and the laser of scheme A can be re-used. The laser beam for the first step is generated using a dye laser ($\sim\,2\pi\times500\,\text{kHz}$ linewidth) pumped by a frequency-doubled Nd:YAG laser. The 553\,nm beam is focused to a waist of 230\,\micro m, overlapped with the 413\,nm laser and propagates on axis through the trap.

Neutral barium atoms are provided by a thermal beam evaporated into the trapping region by resistively heating a tantalum tube of 1\,mm inner diameter filled with barium. A heating current of 3.15\,A results in an oven temperature at thermal equilibrium of approximately 600\,K. To minimize Doppler broadening and to avoid overall Doppler shifts, a collimated atomic beam perpendicular to the photoionization lasers should be used, especially when a high grade of isotope selectivity is required. However, in the current setup, the atomic barium beam encloses an angle of approximately $100\pm5^\circ$ with the direction of laser propagation and the transition frequency is thus red-shifted. The atomic beam is collimated by a 200\,\micro m wide slit aperture at a distance of 4\,cm from the trap axis in the direction perpendicular to the laser beam to ensure a minimal contamination of the trap electrodes with barium.

\begin{figure}
	\centering
		\includegraphics[width=0.48\textwidth]{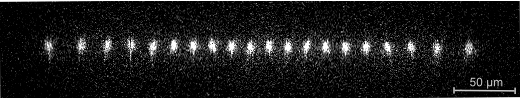}
	\caption{Fluorescence image of 20 \Ba ions photoionized via a dipole-allowed transition (scheme B) and confined in the Paul trap. The ions are sufficiently laser cooled to form an ion Coulomb crystal. The number of confined particles and the trap parameters lead to a linear chain along the axis of the linear trap. A sufficiently large isotope shift of the transition relative to $^{138}\text{Ba}$ ($^{137}\text{Ba}$: $2\pi\times215\,\text{MHz}$; $^{136}\text{Ba}$: $2\pi\times128\,\text{MHz}$; $^{135}\text{Ba}$: $2\pi\times259\,\text{MHz}$; $^{134}\text{Ba}$: $2\pi\times143\,\text{MHz}$\cite{Baird79}) in comparison with the observed Doppler- and power-broadened linewidth allows for a high probability for an isotopically pure crystal. An enhancement of the isotope selectivity can be achieved by a detuning towards lower frequency of the first step photoionization laser because $^{138}\text{Ba}$ has the lowest frequency of all stable isotopes. $\textsuperscript{138}\text{Ba}$ has a natural abundance of 71.7\%. The probability to load an isotopically clean ion crystal consisting of 20 \Ba ions (as shown in the figure) with a non-isotope selective technique amounts to $0.717^{20}\approx0.1\%$.}
	\label{fig:Ba_ionen}
\end{figure}

After ionization, the \Ba ions are laser-cooled on the \mbox{$^2\text{S}_{1/2}\rightarrow ^2\text{P}_{1/2}$} transition at 493\,nm. Figure \ref{fig:Ba_energy_level}b) shows the relevant electronic levels and transitions of singly ionized barium. The upper state decays with a branching ratio of approximately 3:1 into the ground and a metastable $^2\text{D}_{3/2}$ state\cite{davidson92}. The population in the D state is re-pumped to the P state by an additional laser at 650\,nm to approach a closed cycling transition for laser cooling. In addition, a weak magnetic field (0.5\,mT) perpendicular to the linear polarization of the laser beam is used to prevent dark states in the Zeeman manifold of the metastable state. The 493\,nm light is produced by frequency-doubling a commercial narrow-linwidth (typically $2\pi\times1\,\text{MHz}$), tunable diode laser operating in Littrow configuration. Non-critical phase matching in a 10\,mm-long $\text{KNbO}_3$ non-linear crystal is used to double the frequency of the light in a home-build second-harmonic generation external ring cavity\cite{friedenauer06}. The re-pumper light is directly produced by an external cavity diode laser. Both, cooling laser and repumper are overlapped and directed along the axis of the linear trap. Figure \ref{fig:Ba_ionen} shows a two-dimensional fluorescence image of a pure \Ba crystal containing 20 ions loaded into the trap via scheme B. The crystalline structure allows to accurately determine the number of ions of the specific barium isotope that is resonant with the Doppler cooling light. Throughout this paper loading rates of ions are determined by the evaluation of fluorescence images of ions confined in the Paul trap after a certain loading duration. The position and number of non-fluorescing ions within the crystalline structure, for example other barium isotopes or molecular ions can be determined by exploiting the inherent symmetry of Coulomb crystals.

\section{Results}
\label{results}
\begin{figure} 
	\centering
		\subfigure[Scheme A]{\label{Innsbruck}\includegraphics[width=0.50\textwidth]{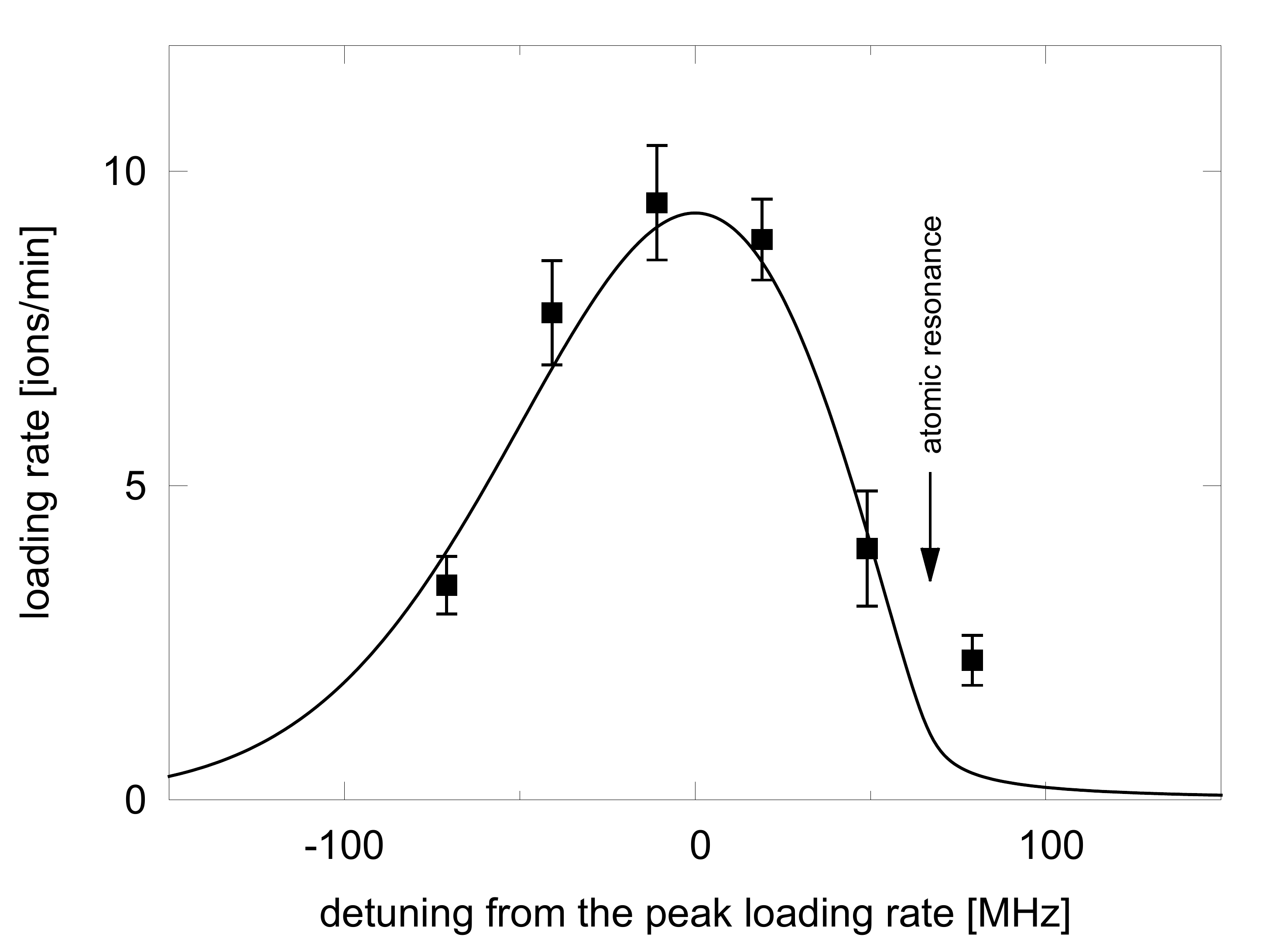}}
		\subfigure[Scheme B]{\label{munich}\includegraphics[width=0.50\textwidth]{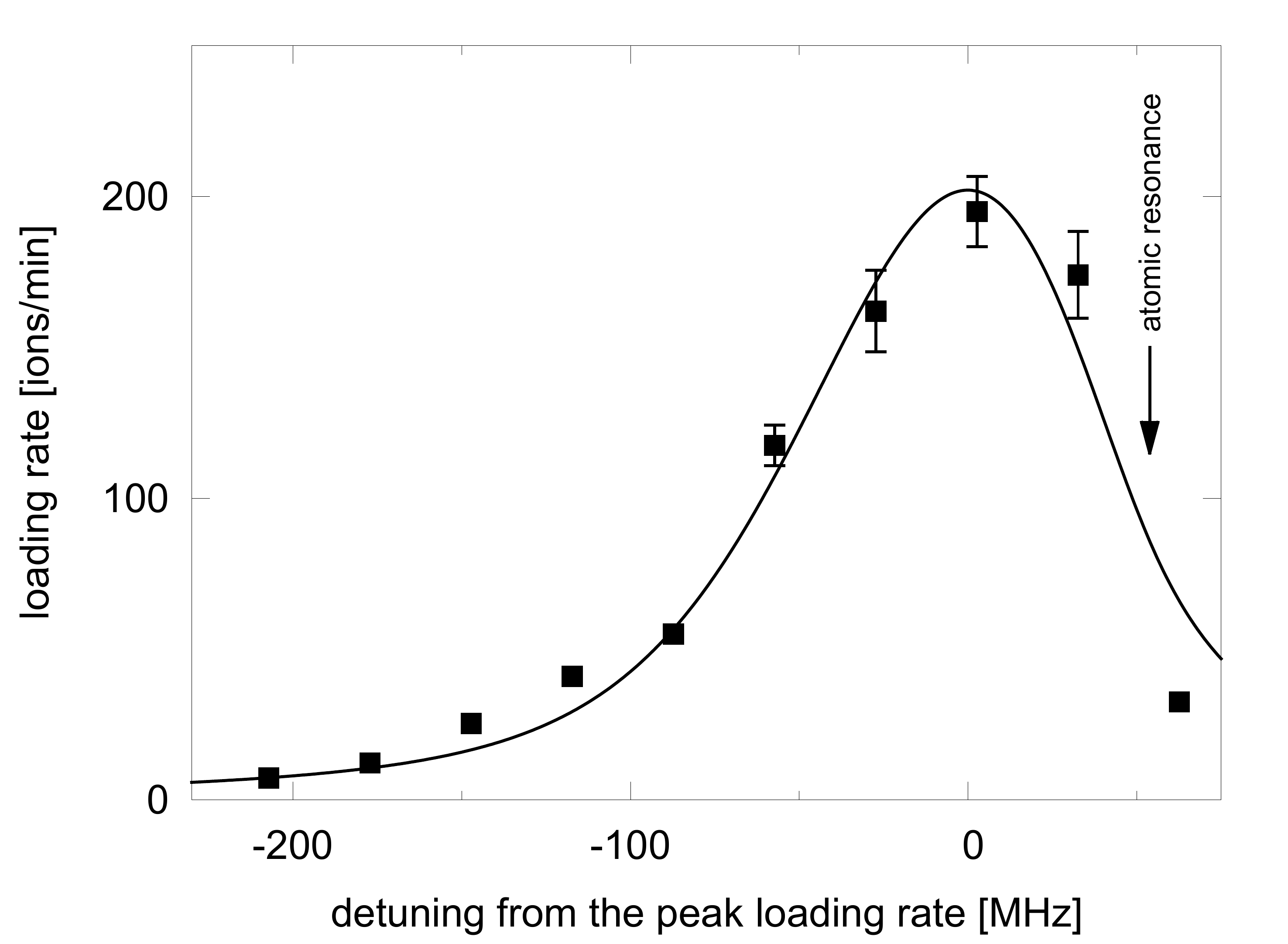}}
	\caption{Loading rate of \Ba ions into the linear Paul trap as a function of the detuning of the laser used in the first ionization step to the peak loading rate. In \ref{Innsbruck}, \ref{munich} the results for scheme A,B are shown respectively. In both cases, the laser powers were set to 1\,mW. In the present setup, the atoms have a velocity component along the propagation direction of the laser and therefore, the peak of the loading rate is red-shifted with respect to the atomic resonance indicated by an arrow. Note the asymmetry of the fitted model curve (black solid lines) in both photoionization schemes which is also due to the non-perpendicular orientation of the laser beam and atomic oven. The given errorbars represent statistical errors only.}
	\label{fig:frequency_scan_innsbruck_scheme}
\end{figure}

In Fig. \ref{fig:frequency_scan_innsbruck_scheme},
the ion loading rates are shown as a function of
the detuning of the laser from the overall shifted atomic resonance (values in this section are given divided by $2\pi$) coupling the ground and the intermediate state (first step laser) for scheme A and B
(the powers of the laser beams providing the first and the second step amount to approximately 1 mW).
The zero of the horizontal axis is calibrated by fitting the maximum of the loading rate.
The solid lines represent the results of a model. Details of the model and discussions about the broadening of the spectral line can be found in section \ref{Theoretical description}. In the present setup, the atoms have a velocity component along the propagation direction of the laser beams, and hence the transition is Doppler-broadened and the maximum loading rate is Doppler-shifted towards lower frequencies. The center of the unshifted atomic resonance is indicated by an arrow
in the figure.

\begin{figure}
	\centering
		\includegraphics[width=0.50\textwidth]{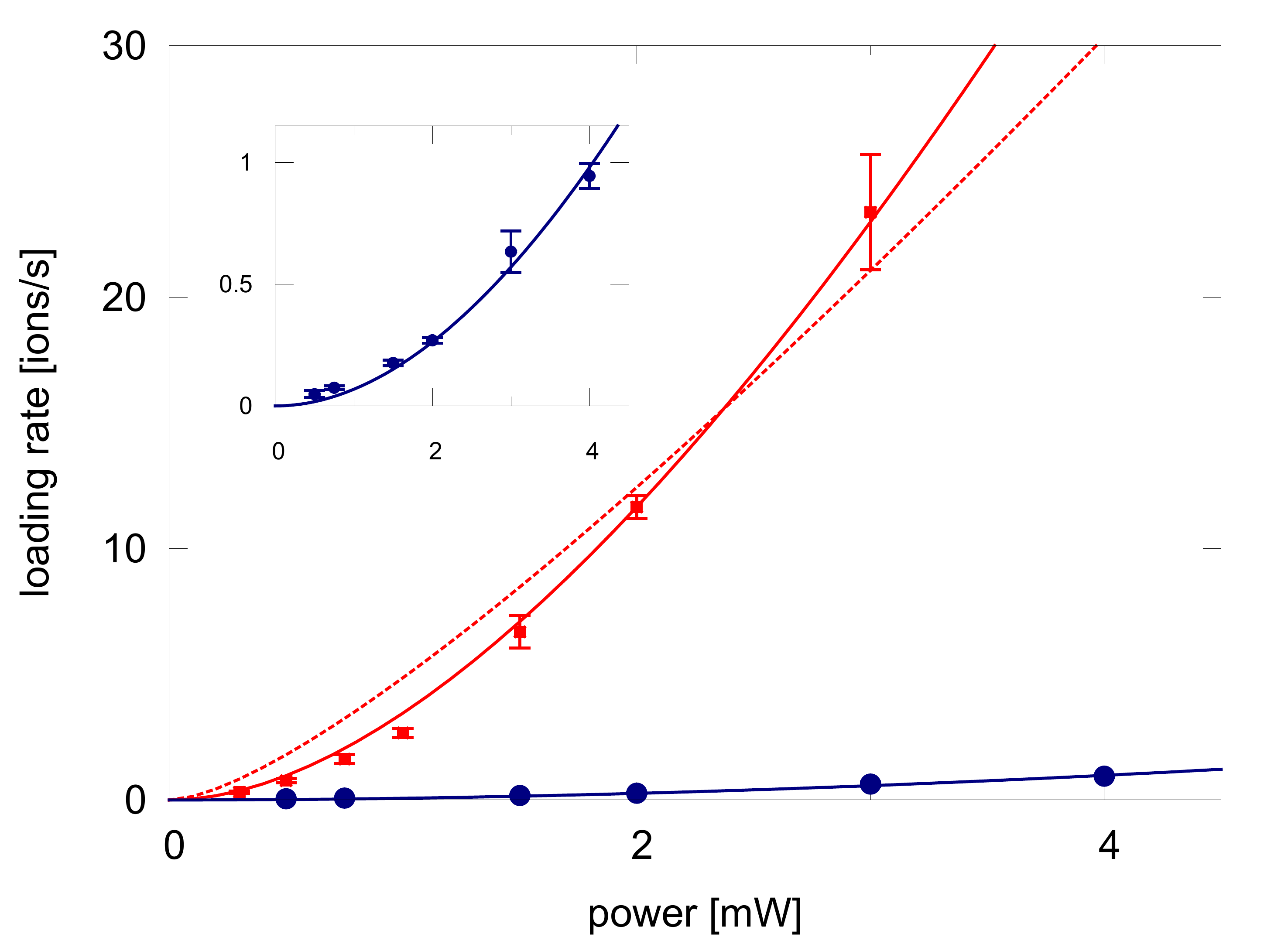}
		\caption{Loading rate achieved with scheme A (blue circles) and B (red squares) as a function of the laser power used in the first ionization step. For scheme B, the power of the 553\,nm and the 413\,nm laser are kept equal. The dashed and solid curves represent the results of a calculation, assuming an atomic oven temperature of T=600\,K. For the dased red curve, the detuning of the first step laser is set to the frequency at which the loading rate is at its maximum and the angle $\theta$ between the laser and the atomic oven is taken to $98.3^\circ$. The remaining fitting parameter is the total efficiency $\alpha$ (see main text). The additional solid red curve shows the result of the fit with the model for scheme B when a detuning of +75\,MHz from the peak loading rate is assumed. This might explain the discrepancy of the results from the model at low powers. The experimental data show that the ion loading rate in the experiment following scheme B is already about 40 times larger than following scheme A. Due to different beam waists used to realize scheme B, an additional increase of the efficiency by a factor of approximately 2.5 has to be considered, leading to a total enhancement of the efficiency by approximately two orders of magnitude compared to scheme A (for details see main text). The inset shows a close-up of the experimental data derived following scheme A and the corresponding curve fitted according to the model with the laser frequency set to the value where the loading rate is maximal.}
	\label{fig:power_scan_munich_vs_Innsbruck}
\end{figure}

The experimentally derived loading rates of scheme A (B) are shown in dependence of the laser power in Fig. \ref{fig:power_scan_munich_vs_Innsbruck} (A: blue circles and B: red squares). The data of scheme A (scheme B) show in principle a dependency on the square of the laser power at 413\,nm (product of the laser powers at 413\,nm and 553\,nm), as to be expected in a process requiring two photons. In scheme B, the laser power of the second laser (413\,nm) is kept equal to that of the first laser (553\,nm) and therefore, the two schemes are comparable in terms of laser power. This is justified, because scheme B does in principle not require a separate laser system at 413\,nm as chosen in the current experiment for simplicity, having the laser source of scheme A available. As shown in \cite{steele07,Wang11}, the transition into the continuum does not have to be of narrow bandwidth (different to the first step in scheme A and B) and can be replaced by an incoherent light source. The solid lines present the results of the calculations based on the model. For the given parameters, the loading rate of scheme B is already about 40 times larger than that of scheme A. As already mentioned in section \ref{experiment}, the beam waists of the two lasers used in scheme B differ considerably at the trap center. Atoms can only be ionized when both required beams overlap, since the lifetime of the excited state is much shorter than the typical transit time of the atom through the beam (see also section \ref{Theoretical description} and the caption of Fig. \ref{fig:power_scan_munich_vs_Innsbruck}). Taking the currently not contributing amount of laser power within the 413\,nm beam for scheme B into account, a lower bound for the additionally increase in ionization efficiency can be derived. For the comparison of the two schemes, a homogeneous atom density in the trap volume and an equal probability to trap an ionized atom is assumed. Implying a flat top profile for both beams of different waists would lead to an underestimation of scheme B in the comparison of efficiencies by a factor of $(w_\text{553}/w_\text{413})^2\approx\, 4$. Considering, more realistically, gaussian beams and assuming a concentric overlap, the measured ionization rate achieved via scheme B has to be multiplied at least with a factor of 2.5 for comparison with scheme A. This allows to derive an increase of the total efficiency via scheme B by two orders of magnitude.

In Fig. \ref{fig:power_scan_munich_fixed413}, the dependence of the loading rate on the power of the first laser (553 nm) is shown when the power of the second laser is kept again constant at 1\,mW. Taking saturation effects into account allows to explain the deviation from a linear dependency of the laser power at 553\,nm. The power dependence is similar to that described in reference \cite{steele07} (Fig.6) and follows in principle a dependency on the square root of the power, as to be expected for an increased saturation of the transition. The solid lines represent the results of the calculations based on the model and will be explained in detail in the next section.

\section{Model}
\label{Theoretical description}
\begin{figure}
	\centering
		\includegraphics[width=0.50\textwidth]{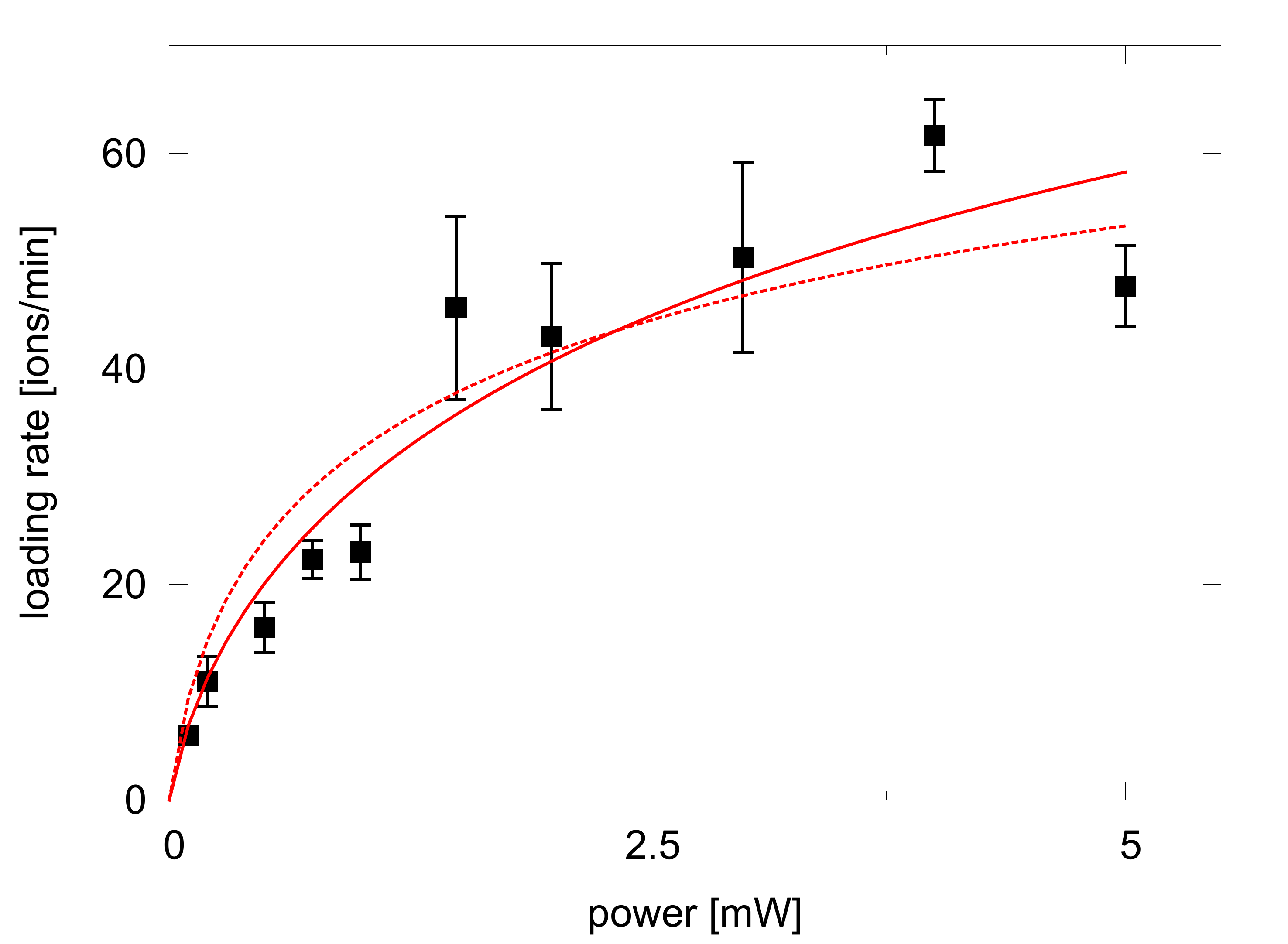}
	\caption{Loading rate achieved via scheme B as a function of the laser power used in the first ionization step. The power of the laser used by the second step is kept at 1\,mW, and the temperature of the atomic oven is assumed to be 550\,K for the calculation, to account for a reduced heating current of the atomic oven compared to the one chosen to obtain the results presented in Figs. \ref{fig:frequency_scan_innsbruck_scheme} and \ref{fig:power_scan_munich_vs_Innsbruck}. The red dashed curve is the result of fitting the model with parameter $\alpha$ and the frequency set at the maximum loading rate. The red solid line is the result of the fit when a detuning of -50\,MHz from this frequency is assumed (see main text).}
	\label{fig:power_scan_munich_fixed413}
\end{figure}

In the present setup, the Doppler broadening of the spectral lines is caused by the temperature $T$ of the atomic ensemble in combination with the angle $\theta$ between the atomic and the laser beam. The derived spectral line of scheme A (Fig. \ref{Innsbruck}) has a full-width-at-half-maximum (FWHM) of 110 MHz. If this effect was caused by Doppler broadening at the given angle $\theta$, the FWHM achieved via scheme B should amount to 82 MHz. The experimental result, however, shows an even larger width (Fig. \ref{munich}). This effect is caused by power broadening, because the transition dipole moment $\mu$ exploited in scheme B is much larger than in scheme A. This is in agreement with the non-linear dependency of the loading rate depicted as a function of laser power in Fig. \ref{fig:power_scan_munich_fixed413}. Here a model is introduced in order to describe and explain the broadening of the spectral lines and the power dependence of the ion loading rate for scheme B with a set of three parameters ($\theta$, $T$, detuning from the resonance frequency of an atom at rest) that provides a consistent description of the experimental situation. A similar model for scheme A will be introduced afterwards.

For convenience, Cartesian coordinates $(x,y,z)$ are used. The laser beams and the atomic oven are placed in the $x-z$ plane. The origin of the coordinate system is chosen at the atomic oven, and the center of the laser beams is at $x=X_0$ and $y=0$, extended along the $z$-axis. The ion loading rate $\eta$ at a specific position $(x,y,z)$ is assumed to be proportional to the atom population at the intermediate state ($6s6p\,^1\text{P}_1$) and to the intensity of the second laser beam transferring the population from the intermediate state to the continuum. The population of the intermediate state $\rho$ depends on the intensity of the first laser beam driving the transition from the ground to the intermediate state.

The transverse intensity distribution of the laser beam is assumed to be gaussian: $I_{i}e^{-2\left[\left(x-X_0\right)^2+y^2\right]/w_i^2}$
($i=1$ (2) for the first (second) step laser, $I_i$ represents the intensity at the center of the laser beam with a beam waist $w_i$). The ionization probability for an atom leaving the oven at $t=0$
with velocity $(v_x,v_y,v_z)$ is given by

\begin{eqnarray}
\label{eq5-1}
\eta(v_x,v_y,v_z)&=&
\alpha \int_0^\infty I_2 e^\Delta \rho\left(v_xt-X_0,v_yt,v_zt\right)dt,\\
\nonumber
\Delta&=&-\frac{2\left[\left(v_xt-X_0\right)^2+
\left(v_yt\right)^2\right]}
{w_2^2}
\end{eqnarray}

where $\alpha$ is a constant of proportionality (see also caption of Fig. \ref{fig:power_scan_munich_vs_Innsbruck}).
The peak intensity is related to the laser power $P_i$ by $I_i=2P_i/\left(\pi w_i^2\right)$.

The population $\rho$ of an atom in the intermediate state is assumed to be stationary. The lifetime in the intermediate state (inverse of the natural linewidth $\Gamma$) is much shorter than the typical transit time of the atom across the laser beam. For example, assuming a Maxwell-Boltzmann distribution for $T=600$ K, the most probable speed of Ba atoms is about 200 m$/$s, and the transit time through the laser beam ($2w_1=460\mu$m) amounts to 2.3 $\mu$s, which is much longer than the lifetime of the $6s6p\,^1\text{P}_1$ state (8\,ns). Therefore, $\rho$ can be expressed as\cite{metcalf99},

\begin{equation}
\rho=
\frac{1}{4}\frac{\Omega_0\left(v_xt,v_yt\right)^2}{\left(\delta
-kv_z\right)^2+\frac{\Gamma^2}{4}+\frac{\Omega_0\left(v_x,v_y
\right)^2}{2}},
\label{eq5-2}
\end{equation}
where $\delta$ is the detuning of the laser from the atomic resonance frequency at rest, $k$ is the wavenumber of the laser beam driving the transition at the Rabi frequency $\Omega_0$ which is a function of the spatial coordinates:

\begin{equation}
\Omega_0\left(x,y\right)^2=\frac{2\mu^2}{\hbar^2\varepsilon_0c}I_{1}
\exp\left\{-\frac{2}{w_1^2}\left[\left(x-X_0\right)^2+y^2\right]\right\}.
\label{eq5-3}
\end{equation}

In Eq. (\ref{eq5-3}), $\hbar$ is the reduced Planck constant, $\varepsilon_0$ is the electric constant and $c$ is the speed of light in vacuum.

The total ion loading rate is obtained by integrating over the atomic velocity for an appropriate velocity distribution function $f(v_r,v_y)$:
\begin{eqnarray}
\eta_0&=&\int_0^\infty v_rdv_r \int_{-\infty}^{\infty}dv_y \int_{\theta-\delta\theta}^{\theta+\delta\theta}d\theta
f(v_r,v_y)\cdot\nonumber\\
&&\eta(v_r\sin\theta,v_y,v_r\cos\theta).
\label{eq5-4}
\end{eqnarray}
Here the coordinates ($r$ and $\theta$), defined as $x=r\sin\theta$ and $z=r\cos\theta$, are introduced and $v_x$ and $v_z$ are replaced by $v_r\sin\theta$ and $v_r\cos\theta$, respectively. The azimuthal angle $\theta$ is the angle included by the direction of the atomic and the laser beam. For a thermal beam of atoms with mass $M$, a Maxwell-Boltzmann distribution can be assumed:

\begin{equation}
f(v_r,v_y)=\frac{1}{N}\exp\left(-M\frac{v_r^2+v_y^2}{2k_BT}\right),
\label{eq5-5}
\end{equation}

where $k_B$ is the Bolzmann constant, and $N$ is a normalization factor.
In the present experimental setup, the slit aperture collimates the beam in the y-direction. However, the diverging angle of the atomic beam in the x-z-plane is negligible compared to the velocity distribution of atoms in a thermal beam. Therefore, the divergence $\delta\theta$ in Eq. (\ref{eq5-4}) approaches zero and $\theta$ is assumed to be at one value in the following.

In the case of scheme A the spontaneous emission from the intermediate state
($5d6p\, ^3\text{D}_1$) mainly leads to metastable states such as $6s5d\, ^3\text{D}_{1,2}$.
The maximal spontaneous emission rate from $5d6p\, ^3\text{D}_1$ to
the ground state ($6s^2\,^1\text{S}_0$) amounts to $1.5\times 10^6\,\text{s}^{-1}$,
whereas that to $6s5d\, ^3\text{D}_1$ is $3.8\times 10^7\,\text{s}^{-1}$ and
that to $6s5d\, ^3\text{D}_2$ is $1.9\times10^7\,\text{s}^{-1}$.
Therefore, the atoms are optically pumped to the metastable states, and the stationary value of $\rho$ approaches zero.
By changing the definition of $\rho$ to be the transition probability from the ground state
to the intermediate state, it is possible to use the same formulae as in Eqs. (\ref{eq5-1}-\ref{eq5-5}),
except for the value of the constant of proportionality $\alpha$.
Thus, applying the present model does not allow for comparing the absolute value of the loading rates achieved via the two schemes.

By substituting Eqs. (\ref{eq5-1}-\ref{eq5-3}) and (\ref{eq5-5}) into
Eq. (\ref{eq5-4}),
the theoretical curves in Fig. \ref{fig:frequency_scan_innsbruck_scheme} are obtained.
In the case of scheme A, the temperature of the atomic beam is assumed to be T=600\,K,
the spontaneous emission rate is $5.85\times10^7\,\text{s}^{-1}$,
and the parameters of the laser providing the first and the second step of ionization are identical. For scheme B, the temperature is also set to T=600\,K, but the spontaneous emission rate is
$1.2\times 10^8\,\text{s}^{-1}$, and the parameters of the first and second step lasers
are set independently. In the calculation, the parameters $\theta$, $\alpha$, and the center frequency of the resonance
are left as fitting parameters for the data presented in Fig. \ref{Innsbruck} and $\theta=98.3$ degrees is deduced. The center of the resonance is pointed at by an arrow in the figure. In Fig. \ref{munich}, $\theta$ is fixed to the obtained value above, and only the center frequency of the
resonance and $\alpha$ remain as fitting parameters.

Using the same model, the results of the derivation for the dependency of the ion loading rate on the power is also
depicted in Figs. \ref{fig:power_scan_munich_vs_Innsbruck} and \ref{fig:power_scan_munich_fixed413}.
Here the only fitting parameter is $\alpha$,
and the detuning is set to the frequency at which the loading
rate is at its maximum. In Fig. \ref{fig:power_scan_munich_fixed413},
however, the red dashed curve derived from the model does not fit the experimental result at small laser power very well. Scrutinizing the experimental data, it turned out that the laser frequency has drifted a few tens of megahertz to the red of the frequency for the peak loading rate. Assuming a detuning of $-50\,\text{MHz}$ from the frequency of maximal loading rate, the result of the model (red solid line) is in better agreement with the experimental findings.

\section{Conclusions and outlook} 
\label{outlook}
In conclusion, this work demonstrated a novel, isotope-selective resonant two-photon photoionization scheme for barium using a dipole-allowed transition. The loading rates into a linear Paul trap was compared to an established scheme using a weak intercombination line. An increase in efficiency by two orders of magnitude was found. This allows to enhance the advantages of photoionization for barium like minimal charge build-up on insulators compared to electron bombardment ionization or patch potentials caused by a contamination of the trap electrodes. Reducing these effects might have a great impact in experiments that rely on high efficiency such as in cavity quantum-electrodynamics with ion coulomb crystals\cite{Albert11}, one-dimensional surface traps for quantum information processing\cite{Stick06}, two-dimensional trapping arrays for quantum simulations\cite{schatz07,Schneider11} or the recently realized optical trapping of ions\cite{Schneider10}. The direct, efficient photoionization of an optically trapped barium atom reduces detrimental charging effects, followed by a loss of the ion out of the shallow, compared to RF-potentials, optical potential. This might allow, for example, to study cold-chemistry processes\cite{Krych11} and to simultaneously avoid micro-motion that normally occurs in conventional ion traps\cite{Cormick11}. In addition, providing loading rates of $10^4$ ions/s are required to allow for experiments at repetition rates of the order of 0.1-1\,kHz, for example, exploring controlled molecular ions that are sympathetically cooled via Ba ions. The devices combining the species, a storage ring\cite{Drewsen2008,schramm01,schramm02} or a molecular conveyor belt\cite{mghpapier}, will substantially depend on the sufficient efficiency of the Ba-ion source.
 
The presented technique can be further advanced by exploiting the tuneability of the 413\,nm laser and its proximity to the ionization threshold. An increase in efficiency is expected by tuning the laser on resonance with a transition to either a field ionized Rydberg state\cite{Gulde01} or with an autoionization resonance\cite{Hudson70,Leeuwen94}. The presented technique can be simplified because the dye laser used to generate the 553\,nm light can be substituted by a low maintenance, all-solid-state laser system. Very recently, a commercial frequency-doubled diode laser system with the required wavelength and power became available. Since the second ionization step does not require a laser of narrow linewidth, substituting the 413\,nm laser by an incoherent light source is possible\cite{steele07,Wang11} and the presented highly efficient photoionization scheme could thus be implemented with modest effort and a robust setup.

\begin{acknowledgements}
Financial support is gratefully acknowledged by the Deutsche Forschungsgemeinschaft, the DFG cluster of Excellence: Munich Center for Advanced Photonics, the International Max Planck Research School on Advanced Photon Science (IMPRS-APS) and the EU research project PICC: The Physics of Ion Coulomb Crystals, funded under the European Community's 7th Framework Programme. The authors would like to thank S. Kahra for his contributions to the barium trapping setup and M. Albert for carefully reading the manuscript. We would also like to thank J. Bayerl, C. Kerzl and T. Dou for their contributions and the Quantum Optics and Spectroscopy Group in Innsbruck especially R. Blatt and D. Rotter for helpful discussions and insights.
\end{acknowledgements}
\newpage

\bibliography{referenzen}   

\begin{thebibliography}{10}

\bibitem{rosenband08}
Rosenband, T., Hume, D.~B., Schmidt, P.~O., Chou, C.~W., Brusch, A., Lorini,
  L., Oskay, W.~H., Drullinger, R.~E., Fortier, T.~M., Stalnaker, J.~E.,
  Diddams, S.~A., Swann, W.~C., Newbury, N.~R., Itano, W.~M., Wineland, D.~J.,
  and Bergquist, J.~C.
\newblock {\em Science}{ \bf 319}(5871), 1808--1812 3  (2008).

\bibitem{ChouScience10}
Chou, C.~W., Hume, D.~B., Rosenband, T., and Wineland, D.~J.
\newblock {\em Science}{ \bf 329}(5999), 1630--1633 (2010).

\bibitem{chou10}
Chou, C.-w., Hume, D.~B., Koelemeij, J. C.~J., Wineland, D.~J., and Rosenband,
  T.
\newblock {\em Physical Review Letters}{ \bf 104}(7), 070802 2  (2010).

\bibitem{haeffner08}
Haeffner, H., Roos, C.~F., and Blatt, R.
\newblock {\em Physics Reports}{ \bf 469}, 155 (2008).

\bibitem{Home09}
Home, J.~P., Hanneke, D., Jost, J.~D., Amini, J.~M., Leibfried, D., and
  Wineland, D.~J.
\newblock {\em Science}{ \bf 325}(5945), 1227--1230 (2009).

\bibitem{Friedenauer08}
Friedenauer, A., Schmitz, H., Glueckert, J.~T., Porras, D., and Schaetz, T.
\newblock {\em Nature Physics}{ \bf 4}(10), 757--761 July  (2008).

\bibitem{Lanyon11}
Lanyon, B.~P., Hempel, C., Nigg, D., Müller, M., Gerritsma, R., Zähringer,
  F., Schindler, P., Barreiro, J.~T., Rambach, M., Kirchmair, G., Hennrich, M.,
  Zoller, P., Blatt, R., and Roos, C.~F.
\newblock {\em Science}{ \bf } (2011).

\bibitem{Paul90}
Paul, W.
\newblock {\em Rev. Mod. Phys.}{ \bf 62}, 531--540 Jul  (1990).

\bibitem{Gulde01}
Gulde, S., Rotter, D., Barton, P., Schmidt-Kaler, F., Blatt, R., and
  Hogervorst, W.
\newblock {\em Applied Physics B: Lasers and Optics}{ \bf 73}, 861--86tur
  (2001).
\newblock 10.1007/s003400100749.

\bibitem{Turchette00}
Turchette, Q.~A., Kielpinski, King, B.~E., Leibfried, D., Meekhof, D.~M.,
  Myatt, C.~J., Rowe, M.~A., Sackett, C.~A., Wood, C.~S., Itano, W.~M., Monroe,
  C., and Wineland, D.~J.
\newblock {\em Phys. Rev. A}{ \bf 61}, 063418 May  (2000).

\bibitem{Stick06}
Stick, D., H. W. O. S. M. M. S. K. M.~C.
\newblock {\em Nature Physics}{ \bf 2}(1), 36--39 (2006).

\bibitem{Hurst79}
Hurst, G.~S., Payne, M.~G., Kramer, S.~D., and Young, J.~P.
\newblock {\em Rev. Mod. Phys.}{ \bf 51}, 767--819 Oct  (1979).

\bibitem{steele07}
Steele, A.~V., Churchill, L.~R., Griffin, P.~F., and Chapman, M.~S.
\newblock {\em Physical Review A}{ \bf 75}(5), 053404 5  (2007).

\bibitem{Wang11}
Wang, B., Zhang, J.~W., Gao, C., and Wang, L.~J.
\newblock {\em Opt. Express}{ \bf 19}(17), 16438--16447 Aug  (2011).

\bibitem{kjaegaard00}
Kjaergaard, N., Hornekaer, L., Thommesen, A., Videsen, Z., and Drewsen, M.
\newblock {\em Applied Physics B: Lasers and Optics}{ \bf 71}(2), 207--210
  August  (2000).

\bibitem{lucas04}
Lucas, D.~M., Ramos, A., Home, J.~P., McDonnell, M.~J., Nakayama, S., Stacey,
  J.-P., Webster, S.~C., Stacey, D.~N., and Steane, A.~M.
\newblock {\em Phys. Rev. A}{ \bf 69}(1), 012711 Jan  (2004).

\bibitem{Tanaka05}
Tanaka, U., Matsunishi, H., Morita, I., and Urabe, S.
\newblock {\em Applied Physics B: Lasers and Optics}{ \bf 81}, 795--799 (2005).
\newblock 10.1007/s00340-005-1967-2.

\bibitem{Johanning11}
Johanning, M., Braun, A., Eiteneuer, D., Paape, C., Balzer, C., Neuhauser, W.,
  and Wunderlich, C.
\newblock {\em Applied Physics B: Lasers and Optics}{ \bf 103}, 327--338
  (2011).
\newblock 10.1007/s00340-011-4502-7.

\bibitem{Brownnutt07}
Brownnutt, M., Letchumanan, V., Wilpers, G., Thompson, R., Gill, P., and
  Sinclair, A.
\newblock {\em Applied Physics B: Lasers and Optics}{ \bf 87}, 411--415 (2007).
\newblock 10.1007/s00340-007-2624-8.

\bibitem{vant06}
Vant, K., Chiaverini, J., Lybarger, W., and Berkeland, D.~J.
\newblock  (2006).

\bibitem{phd_ludsteck}
Ludsteck, V.
\newblock {\em Experimente mit einer linearen Ionenkette zur Realisierung eines
  Quantencomputers}.
\newblock PhD thesis, Fakultät fuer Physik, Ludwig-Maximilians-Universtitaet
  Muenchen,  May  (2004).

\bibitem{Hendricks07}
Hendricks, R., Grant, D., Herskind, P., Dantan, A., and Drewsen, M.
\newblock {\em Applied Physics B: Lasers and Optics}{ \bf 88}, 507--513 (2007).
\newblock 10.1007/s00340-007-2698-3.

\bibitem{cetina07}
Cetina, M., Grier, A., Campbell, J., Chuang, I., and
  Vuleti\ifmmode~\acute{c}\else \'{c}\fi{}, V.
\newblock {\em Phys. Rev. A}{ \bf 76}(4), 041401 Oct  (2007).

\bibitem{deslauriersmonroe06}
Deslauriers, L., Acton, M., Blinov, B.~B., Brickman, K.-A., Haljan, P.~C.,
  Hensinger, W.~K., Hucul, D., Katnik, S., Kohn, R.~N., Lee, P.~J., Madsen,
  M.~J., Maunz, P., Olmschenk, S., Moehring, D.~L., Stick, D., Sterk, J., Yeo,
  M., Younge, K.~C., and Monroe, C.
\newblock {\em Phys. Rev. A}{ \bf 74}(6), 063421 Dec  (2006).

\bibitem{leibrandt07}
Leibrandt, D.~R., Clark, R.~J., Labaziewicz, J., Antohi, P., Bakr, W., Brown,
  K.~R., and Chuang, I.~L.
\newblock {\em Phys. Rev. A}{ \bf 76}(5), 055403 Nov  (2007).

\bibitem{rotter08}
Rotter, D.
\newblock {\em Quantum feedback and quantum correlation measurements with a
  single Barium ion}.
\newblock PhD thesis,  (2008).

\bibitem{diedrich87b}
Diedrich, F., Peik, E., Chen, J.~M., Quint, W., and Walther, H.
\newblock {\em Physical Review Letters}{ \bf 59}(26), 2931 12  (1987).

\bibitem{Wineland87}
Wineland, D.~J., Bergquist, J.~C., Itano, W.~M., Bollinger, J.~J., and Manney,
  C.~H.
\newblock {\em Physical Review Letters}{ \bf 59}(26), 2935 12  (1987).

\bibitem{Armstrong93}
Armstrong, D.~J. and Cooper, J.
\newblock {\em Phys. Rev. A}{ \bf 47}, R2446--R2449 Apr  (1993).

\bibitem{davidson92}
Davidson, M.~D., Snoek, L.~C., Volten, H., and Doenszelmann, A.
\newblock {\em Astronomy And Astrophysics}{ \bf 255}(1-2), 457--458 February
  (1992).

\bibitem{friedenauer06}
Friedenauer, A., Markert, F., Schmitz, H., Petersen, L., Kahra, S., Herrmann,
  M., Udem, T., Haensch, T., and Schaetz, T.
\newblock {\em Applied Physics B: Lasers and Optics}{ \bf 84}(3), 371--373
  January  (2006).

\bibitem{Baird79}
Baird, P. E.~G., Brambley, R.~J., Burnett, K., Stacey, D.~N., Warrington,
  D.~M., and Woodgate, G.~K.
\newblock {\em Proceedings of the Royal Society of London. Series A,
  Mathematical and Physical Sciences}{ \bf 365}(1723), pp. 567--582 (1979).

\bibitem{metcalf99}
Metcalf, H.~J. and Van~der Straten, P.
\newblock {\em Laser cooling and trapping}.
\newblock Springer, New York , London,  (1999).

\bibitem{Albert11}
Albert, M., Dantan, A., and Drewsen, M.
\newblock {\em Nat Photon}{ \bf advance online publication}, 2011/09/04/online
  (2011).

\bibitem{schatz07}
Schaetz, T., Friedenauer, A., Schmitz, H., Petersen, L., and Kahra, S.
\newblock {\em J. Mod. Opt.}{ \bf 54}, 2317--2325 Nov  (2007).

\bibitem{Schneider11}
Schneider, C., Porras, D., and Schaetz, T.
\newblock {\em arXiv:1106.2597v1}{ \bf } (2011).

\bibitem{Schneider10}
Schneider, C., Enderlein, M., Huber, T., and Schaetz, T.
\newblock {\em Nature Photonics}{ \bf 4}(October), 5 (2010).

\bibitem{Krych11}
Krych, M., Skomorowski, W., Paw\l{}owski, F., Moszynski, R., and Idziaszek, Z.
\newblock {\em Phys. Rev. A}{ \bf 83}, 032723 Mar  (2011).

\bibitem{Cormick11}
Cormick, C., Schaetz, T., and Morigi, G.
\newblock {\em New Journal of Physics}{ \bf 13}(4), 043019 (2011).

\bibitem{Drewsen2008}
http://www.xfel.eu/sites/site-xfel-gmbh/content/e63594
  /e63599/e81234/e65128/e76474/sqs-wg-ii-report-eng.pdf.

\bibitem{schramm01}
Schramm, U., Schaetz, T., and Habs, D.
\newblock {\em Phys. Rev. Lett.}{ \bf 87}, 184801 Oct  (2001).

\bibitem{schramm02}
Schramm, U., Schaetz, T., and Habs, D.
\newblock {\em Phys. Rev. E}{ \bf 66}, 036501 Sep  (2002).

\bibitem{mghpapier}
Kahra, S. and Leschhorn, G.
\newblock Controlled delivery of single molecules into ultra-short laser
  pulses: a molecular conveyor belt, to be published.

\bibitem{Hudson70}
Hudson, R.~D., Carter, V.~L., and Young, P.~A.
\newblock {\em Phys. Rev. A}{ \bf 2}, 643--648 Sep  (1970).

\bibitem{Leeuwen94}
van Leeuwen, R., Ubachs, W., and Hogervorst, W.
\newblock {\em Journal of Physics B: Atomic, Molecular and Optical Physics}{
  \bf 27}(17), 3891 (1994).

\end{thebibliography}
\bibliographystyle{nature}

\end{document}